\documentstyle[aps]{revtex}
\begin{document}

\title{On Unitarity Based Relations Between Various \\Lepton Family Violating
Processes}

\author{S. Nussinov}
\address{\it Department of Physics,Tel-Aviv University, Ramat-Aviv, Tel-Aviv, Israel}
\author{R. D. Peccei}
\address{\it Department of Physics and Astronomy, University of California, Los Angeles, CA 90095-1547} 
\author{X. M. Zhang}
\address{\it Institute of High Energy Physics, Beijing 100039, PR China}

\maketitle

\begin{abstract}
Simple ``unitarity inspired" relations between two- and
three-body lepton flavor violating decays are noted and discussed. In the absence of cancellations, the existing strong bounds on $\mu \to 3e$ and $ \mu\to e\gamma\gamma$ severly constrain two-body lepton flavor violating decays.
\end{abstract}

Lepton flavor violating (LFV) processes are strongly suppressed in the standard
model by powers of (small) neutrino masses.  Such decays signal therefore
new physics.  At present we have stringent bounds for $\mu$ decays [BR$(\mu\to 3e)\leq
10^{-12}$, BR$(\mu\to e\gamma\gamma) \leq 10^{-10}$] 
 and somewhat weaker
$O(10^{-6})$ bounds on LFV $\tau$ decays.~\cite{PDG}

The availability of large samples of decaying vector bosons [$V=J/\psi,~\Upsilon$, and
$Z^o$] or pseudoscalars [$\pi^o,\eta$] and the clear signature provided by
$\mu^\pm e^\mp$ final states suggests searching for LFV two-body decays
$V\to \mu^\pm e^\mp$ or $\pi^o/\eta\to \mu^\pm e^\mp$.  In this note we show that rather simple considerations, based on unitarity, provide rather strong constraints on two-body LFV processes. Hence, most three-body $\mu$ and $\tau$ LFV decays are likely to
remain more sensitive tests of lepton flavor violation, rather than the corresponding two-body decays.

\section*{Basic Considerations}

Let us assume that a vector boson $V_i$ [ Here $V_i$ could be either a  fundamental state, like the $Z^o$,
or a quark-antiquark bound state like the $\phi,J/\psi$, or $\Upsilon$] couples to $\mu^\pm e^\mp$.  If it couples also
to $e^+e^-$--- as all the states above do--- then by unitarity its exchange contributes
also to $\mu\to 3e$.  Let us write the effective coupling between the vector boson $V_i$ and $\mu^\pm e^\mp$ as
\begin{equation}
{\cal L}_{\rm eff}=\tilde g_{V\mu e}\bar\mu\gamma_\alpha e
V^\alpha ~+{\rm h. c.}~.
\end{equation}
This coupling, through the diagram of Fig. 1, contributes to the $A(\mu\to 3e)$
amplitude a term 
\begin{equation}
A(\mu \to e)=\bar u_\mu(p)\gamma^\alpha u_e(k_3)\bar v_e(k_1)\gamma_\alpha
u_e(k_2)\cdot\frac{ \tilde{g}_{V\mu e}g_{Vee}}{M_V^2-s}~. 
\end{equation}
Here $g_{Vee}$ is the effective coupling of the vector boson $V_i$ to $e^+e^-$, while $s\equiv (k_1+k_2)^2\leq m_\mu^2$. ~\footnote{There are, of course, also axial vector couplings of the $Z^o$ to $e^+e^-$, which contribute to this process. These have not been included in the above, since they do not change our qualitative discussion. These couplings are, however, taken into account in the $Z$ bounds given in Eqs. (5) and (9) below.}
As a first approximation, it is sensible to neglect $s$ in comparison
with $M_V^2$. Then comparing the above contribution to the $\mu \to 3e$ process to that of ordinary muon decay, $\mu\to e\nu\bar \nu$, which proceeds via $W$ exchange and (almost) identical kinematics, gives the relation:
\begin{equation}
\frac{[\Gamma(\mu\to 3e)]_{V-{\rm exch.}}}{\Gamma(\mu\to e\nu\bar \nu)}\approx
\frac{\tilde g_{V\mu e}^2g_{Vee}^2}{M_V^4}/\frac{g_W^4}{M_W^4}~.
\end{equation}
Since $\Gamma(V\to e^+e^-)\sim g_{Vee}^2M_V$ and $\Gamma(V\to \mu e) \sim \tilde g_{V\mu e}^2 M_V$, while
$\Gamma(W\to e\nu)\sim g_W^2M_W$,
we can rewrite the last expression as
\begin{equation}
[{\rm BR}(\mu\to 3e)]_{V-{\rm exch.}} \approx 
\frac{\Gamma(V\to e^+e^-)\Gamma(V\to \mu^\pm e^\mp)}{\Gamma^2(W\to e\nu)}
\left(\frac{M_W}{M_V}\right)^6~.
\end{equation}
Using BR$(\mu\to 3e)\leq 10^{-12}$ and other data pertaining to the
$e^+e^-$ widths of the various vector mesons $V_i$, we find a set of bounds for the two-body LFV branching ratios of these vector bosons. These bounds are:
\begin{eqnarray}
{\rm BR}(Z^o\to \mu e) &\leq& 5 \times 10^{-13} \\
{\rm BR}(J/\psi\to\mu e)&\leq&4 \times 10^{-13} \\
{\rm BR}(\Upsilon\to \mu e) &\leq&2 \times 10^{-9} \\
{\rm BR}(\phi\to\mu e) &\leq& 4 \times 10^{-17}~.
\end{eqnarray}
Likewise, the generic upper bounds on LFV tau decays BR$(\tau\to 
\ell \ell^\prime\bar\ell^\prime)\leq 10^{-6}$ yields
\begin{eqnarray}
{\rm BR}(Z^o\to \tau\bar \ell) &\leq& 3 \times 10^{-6} \\
{\rm BR}(J/\psi\to \tau\bar\ell) &\leq&6 \times 10^{-7} \\
{\rm BR}(\Upsilon\to\tau\bar\ell) &\leq& 10^{-2}~,
\end{eqnarray}
with $\ell/\ell^\prime = e/\mu$.  Except for (11),  these inferred bounds
are unlikely to be improved by future experimental data on two-body decays.

One can use similar considerations to obtain bounds on the LFV decays of pseudoscalar states. For these purposes, one considers instead of the $\mu \to 3e$ process the $\mu^-\to e^-\gamma\gamma$ decay, which has a LFV bound BR$(\mu\to e\gamma\gamma)\leq 10^{-10}$. For this latter process the LFV couplings of the $\pi^o/\eta^o$ contribute, due to the exchange of these particles in the
s-channel. We can again
utilize this fact to infer upper bounds on $\pi^o/\eta^o\to \mu^\pm e^\mp$.
The $\pi^o/\eta^o\gamma\gamma$ vertex, because of gauge invariance, involves two derivatives:
\begin{equation}
 {\cal L}_{\rm eff}=\frac{\phi}
{f_{\phi}}F^{\mu \nu}\tilde F_{\mu \nu}~,
\end{equation}
where $\phi=\pi^o,\eta^o$. This derivative coupling, in contrast to the $V\to \mu e$ non-derivative coupling
encountered earlier, kinematically suppresses the off-shell
$\pi^o/\eta^0\to\gamma\gamma$ contribution at $s=(k_1+k_2)^2 \leq m_\mu^2$
relative to what it would be for on-shell $\pi^o/\eta^o$ decay.  Consequently
the analog to Eq. (4) for the present case, 
\begin{equation}
[{\rm BR}(\mu\to e\gamma\gamma)]_{\pi^o/\eta^o-{\rm exch.}}\approx \frac{\Gamma(\pi^o/\eta^o\to\gamma\gamma)
\cdot\Gamma(\pi^o/\eta^o\to \mu^\pm e^\mp)}{\Gamma^2(W\to e\nu)}
\left(\frac{M_W}{m_{\pi^/\eta}}\right)^6\cdot
\left(\frac{\langle s_{\gamma\gamma}\rangle}{m_{\pi/\eta}^2}\right)^2~,
\end{equation}
contains an extra factor
\begin{equation}
\frac{\langle s^2_{\gamma\gamma}\rangle}{m^4_{\pi/\eta}} \approx \left(
\frac{m_\mu}{2m_{\pi/\eta}}\right)^4~,
\end{equation}
which tends to weaken the bounds one can derive. One finds, for pseudoscalar LFV decays the bounds:
\begin{eqnarray}
{\rm BR}(\eta\to\mu e) &\leq& 10^{-8} \\
{\rm BR}(\pi^o\to\mu e) &\leq& 10^{-10}~.
\end{eqnarray}

In the discussion above we have obtained the quoted bounds purely by concentrating on the contribution of the exchanged state in question to the LFV process. One can imagine, however, additional LFV contributions. For example, for the $\mu^-\to e^-\gamma\gamma$ decay, in addition to $\pi^o/\eta^0$ exchange in the
$s$ channel, we have also the contribution of electron exchange in the $t$ and $u$
channels (see Fig. 2).  In this case, however, the stringent bound on the
$\mu\to e\gamma$ vertex coming from experiment [BR ($\mu\to e\gamma)\leq 5\times 10^{-11}$] strongly
suppresses these additional diagrams and causes negligible modifications to the bounds (15), (16). Even in the absence of a strong bound on the $\mu\to e\gamma$ coupling, we would like to note that cancellations between $s$ and
$t$ channel contributions are in general expected to be at best rather
partial.  Unless all particles, both external and exchanged, are spinless any
specific $s$ channel amplitude will have different $\cos\theta_s$ (or
$\cos\theta_t$) dependence, and will contribute to different combinations
of helicity amplitudes than the $t$ channel exchange
contributions.

By the same token, it is clear that cancellations among different angular momentum
states exchanged in the $s$ channel are also impossible.  Indeed, for example, the total decay rate for $\mu \to 3e$ can be expressed as
\begin{eqnarray}
\Gamma(\mu\to 3e) &=& \int^{\mu^2}_0 ds \sum_{\alpha\beta,\gamma\delta}
(s-4m_e^2)^{-1/2}\left[\frac{\Delta(m_\mu^2,s,m_e^2)}{s}\right]^{-1/2} 
\nonumber \\ 
& &{}\times\sum_J
(2J+1) |A^J_{\alpha\beta,\gamma\delta}(s)|^2
\end{eqnarray}
with $\Delta$ the triangular function expressing the initial C.M. momentum in the
$s$ channel, which here is that of  $\mu\bar e_1$ or, equivalently, $ \bar e_3e_2$.  The $A^J_{\alpha\beta,\gamma\delta}(s)$ are the partial waves in the Jacob-Wick
expansion of the various $s$ channel helicity amplitudes.
Note that for the $\mu\to 3e$  case, adding the $t$ channel amplitude amounts
to enforcing the (anti) symmetrization between the $e_3$ and $e_2$ fermions.
Since Fermi statistics does not preclude the vectorial coupling contributing to $\mu\to 3e$ considered
here [c.f. Eq. (2)], no cancellation of $s$ and $t$ contributions should
arise as well.

\section*{ Possible Limitations  on the  Derived LFV Bounds }

Although  we have called the bounds we obtained above unitarity bounds, in the strict sense the inferred bounds are not true unitarity bounds--as would
be the case if the exchanged particle(s) were on mass shell. To illustrate this point, let us recall a
well known example  of a true unitarity bound arising in rare Kaon decays. This is the lower bound for  the BR$(K_L\to\mu^+\mu^-)$  derived from the measured  branching ratio of $K_L\to\gamma\gamma$.  The $K_L\to \gamma\gamma$
process, with an on-shell
$\gamma\gamma$ intermediate state,  contributes to Im $A(K_L\to\mu^+\mu^-)$
via  unitarity since Im $A(K_L\to\mu^+\mu^-)\sim$
$A(K_L\to\gamma\gamma)* A(\gamma\gamma\to\mu^+\mu^-)$. This contribution provides a strict lower bound to the  BR$(K_L\to\mu^+\mu^-)$, so the apparent violation of this
bound in early $K_L\to\mu^+\mu^-$ data was a source for much concern. Modern day data, as expected, agrees with this bound.~\cite{Kmumu}

In the present context, an example of a ``pure" unitarity bound for a LFV 
process is provided by the ``$\tau$ analog" of the $\phi\to\mu e$ process.  Because the decay
$\phi\to\tau\bar \ell$ is kinematically forbidden, what one should consider
instead is $\tau\to\phi\mu$ (or $\tau\to\phi e$).  The ``on-shell" $\phi$ emitted in this
putative process propagates over a long distance, of order $1/\Gamma_\phi
\simeq 30$ fm, before decaying into $K\bar K,\mu^+\mu^-,e^+e^-$ in a manner
which is completely independent 
of its production.
This will generate a distinct narrow contribution to the corresponding three body processes $\tau\to
K \bar K \mu$, $\tau\to
\mu^+\mu^-\mu $, $\tau\to
e^+e^-\mu$ ,contributing to the imaginary part of these amplitudes at $s=m_\phi^2$.  Hence, for example, there is an attendant lower bound
on BR$[\tau\to \ell\ell^\prime\bar\ell^\prime]$ which is simply
BR$(\tau\to\phi\bar\ell)\cdot{\rm BR}(\phi\to\ell^\prime\bar\ell^\prime)
\simeq 3\times10^{-4}\times {\rm BR}(\tau\to\phi\bar\ell)$.  The resulting
rigorous upper bound one obtains,
\begin{equation}
{\rm BR}(\tau\to\phi\bar\ell)\leq 3\times 10^{-3}~,
\end{equation}
unfortunately happens to be rather weak.  

All the vector (or pseudoscalars) used as
intermediaries in deriving the bounds in Eqs (5)-(11) and Eqs. (15) and (16) are
not on-shell.  Thus we must entertain the possibility that their
contribution to the three-body decays considered are reduced.  This could void, or at least weaken the various strong bounds obtained above.   In the rest of this note we will focus on possible mechanisms for such a reduction.

\subsection*{Kinematical Suppression of the LFV Bounds}

The size of the boson
exchange contribution to the three-body decay amplitude can be reduced if there are {\bf kinematical} suppressions.  These arise when the effective
boson couplings are not minimal, involving derivatives (or momentum
factors).  We already encountered one such case above, when we discussed  the $\pi^o/\eta^o$ contribution to $\mu\to e\gamma\gamma$. We want to discuss here whether such kinematical suppressions may not also affect the vector exchange contributions.

It is clearly possible to imagine that the LFV $V_i\mu^\pm e^\mp$ vertex, instead of having the form of Eq. (1), involves an
anomalous magnetic moment coupling:
\begin{equation}
{\cal L}_{\rm eff}^{\rm Magnetic}=\frac{1}{M_V}\bar\mu
\sigma_{\alpha\beta} e
 (\partial^\alpha V^\beta-\partial^\beta V^\alpha) ~+{\rm h. c.}~.
\end{equation}
In this case
the contribution of the virtual $V_i$ to $\mu\to 3e$ is reduced by
\begin{equation}
\frac{q^2}{M_{V}^2} \approx \frac{\langle s\rangle}{M_{V}^2}\approx
\frac{m_\mu^2}{2M_{V}^2} = \left\{ \begin{array}{lll}
3\times 10^{-3} &    & V=\phi \\
3\times 10^{-4} &    & V=J/\psi \\
3\times 10^{-5} &  & V=\Upsilon \\
3\times 10^{-7} &    & V=Z^o
\end{array} \right\}
\end{equation}
This would considerably weaken the bounds in Eqs. (6)-(8) and reduce the bound on $Z$ decay to only BR$(Z^o\to e\mu)\leq 1.5 \times 10^{-6}$.

It does not seem likely to us, however, that the strong suppression factors appearing in Eq. (20) obtain in practice. Indeed, various model calculation involving mixing among heavy 
neutrinos~\cite{IJR} lead to $Z^o\to e\mu$ and effective $c\bar c\to\mu e$  flavor violating vertices, which involve non anomalous terms---
terms like the $Z\mu e$ coupling  of Eq. (1) and vectorial couplings like
$\bar c\gamma_\alpha c\bar\mu\gamma^\alpha e$. ~\cite{wang}   Hence we believe that the  kinematic suppression given by Eq. (20)  probably should not be included in our bounds.

\subsection*{Dynamical Suppression of the LFV Bounds}

There is another possible source of suppression which needs to be considered. This is connected to  possible  "form factor" effects due  to the {\bf dynamics} which would, for example, reduce the contribution of the
various $V_i$ states to $\mu\to 3e$ compared to the naive expectations. However, the effect of form factors should be minimal or controllable
if the LFV is induced by physics at scales much higher than the EW scale
or  the $Z$ mass.  The effects of dynamics are  nicely illustrated in a recent paper  by Ilana, Jack and Riemann~\cite{IJR}. These authors find, in fact, an apparent mild enhancement when the $Z^o\to\mu e$ process is induced by relatively light
($m_{\nu_i}\leq 45$ GeV) neutrinos. Indeed, in this case the on-shell
contribution of $\nu\bar\nu$ loops enhances the $Z$ decay rates
relative to the $s\approx 0$ contribution by factors of 10-100. However, such light active neutrinos would contribute to the $Z^o$
width and are hence ruled out. Thus such an ehancement is not physically expected. \footnote {The BR considered in Ref. [3] for light neutrinos-- i.e. neutrinos with masses in the eV range, as inferred from the SuperKamiokande~\cite{SuperK} data-- are very much below our bounds}

In terms of the dispersive approach adopted here, such a "form factor"
suppression would result from cancellations in the corresponding partial wave amplitudes. Consider, for example, the $A^{J=1}$ partial wave amplitude for the $\mu \to 3e$ process:

\begin{equation}
A^{J=1}_{\alpha \beta,\gamma \delta}(s) = \sum_i
\frac{g_{V_iee}\tilde g_{V_i\mu e}}{M_i^2-s} +
\int \frac{ds^\prime \rho_{\rm LFV}^{J=1}(s^\prime)}{s^\prime-s}~.
\end{equation}
To get a "form factor" suppression, there must be a cancellation between the  contributions of the various $(V_i)$ particles among themselves, or between these contributions and those of the continuum. Let us examine these possibilities.

For the case of quarkonium intermediate states, besides the lowest energy bound state there are towers of states of the same spin and parity. Thus, for example, in Eq.(21) besides the contribution of the $J/\psi$ one should also take into acount the exchange of the $\psi^\prime,
\psi^{\prime\prime},\ldots,\psi^{(n)}$ charmonium bound resonances. Is it possible that these additional contributions largely cancel the $J/\Psi$ term in Eq. (21)? This is unlikely for the following reason. To get
$J/\psi$ exchange to contribute to $\mu\to 3e$ in the first place, one needs
to assume that the LFV physics at a high scale induces an effective 
four-Fermi coupling of the form: 
\begin{equation}
{\cal L_{\rm eff}}=\tilde G_{c\bar c\mu e}\bar c\gamma_\alpha c\bar\mu\gamma^\alpha e~. 
\end{equation}
Such a coupling underlies all the other charmonium contributions. In fact,
quark-hadron duality identifies the $J/\psi,\psi^\prime,\ldots,\psi^{(n)}$ 
contributions to $\mu \to 3e$ as arising from specific portions of the
$s^\prime \equiv M_{\bar cc}$ integration region where
due to non-perturbative QCD effects $1^{--}$ $c \bar c$ bound states dominate, as shown schematically in Fig. 3.  Both the $g_{\psi^{(n)}ee}$ and $\tilde g_{\psi^{(n)}\mu e}$ couplings appearing in Eq. (21) are proportional to 
the wave function of $\psi^{(n)}$ at the origin, ${\bf {\Psi^{(n)}(0)}}$.
Thus all the terms in the sum share a common sign--- fixed by the sign of $\tilde G_{c\bar c \mu e}\cdot Q_c $, with $Q_c=2/3$ being the charge of the charm quark--- 
and cancellation cannot occur.  Similar arguments apply against possible
cancellations among the various states in the $\Upsilon$ sector.

 The above discussion still leaves open the possibility of cancellations
in the partial wave amplitude between different quarks-antiquark contributions ($c\bar c, b\bar b,s\bar s,\dots$) or, equivalently, between the various resonant states ($J/\psi,\Upsilon,\phi,\ldots$.  While possible this seems highly unlikely.  For example, even if all the effective couplings
$\tilde G_{q_i\bar q_i\mu e}$ were equal due to some universality, and the
bubble kinematics were identical, the net contribution would not vanish since the total contribution would be proportional to
$\sum Q_{q_i}\not= 0$. Furthermore,
for the case of light quarks such cancellations cannot work even in
principle.  The $s$ dependence for $0 \leq s\leq m_\mu^2$ neglected above
implies, for example, that a $\omega/\rho$ and $\phi$ contribution to the total
decay rate cancel only at the level of $(m_\mu^2/3)(m_\phi^2-m_{\rho/\omega}^2)/(m_\phi^2)^2 \approx 10^{-3}$.

\section*{Concluding Remarks}

In general, lepton flavor violating processes have been analyzed within a specific theoretical framework. In this context, the restrictive role played by the low energy bounds ($\mu\to 3e~\mu\to e$ conversion, etc.) has been noted by many authors.~\cite{IJR}~\cite{DT}  In this note instead we
tried to present in a, relatively model-independent manner, the  connections which unitarity implies between some two-body and three-body LFV decays. We have illustrated these connections by focusing on a few processes. Clearly, many other bounds can be obtained. Indeed, since the Particle Data Group~\cite{PDG}
lists altogether about one hundred LFV processes, many additional
results can come from a more comprehensive analysis.  

We have noted that the 
bounds that we derived can be avoided if one can kinematically suppress the small $s$ contributions (e.g. by having a purely anomalous
 magnetic $Z^o\mu e$ coupling), or as a result of some (rather unlikely) cancellations. Because we cannot rule out these possibilities with absolute certainty, we hope that
our discussion will not dissuade future efforts to improve the
bounds on  LFV decays of the $Z,J/\psi, \ldots$.  Such decays would not
only signal new LFV physics but, because of our considerations, this physics must also naturally give cancellations among terms so as to lead to a small $\mu\to 3e$ branching ratio.
Thus searching
for $V_i\to\mu^\pm e^\mp$ decays at levels considerably higher than our suggested
bounds remains a worthwhile experimental challenge. 
\section*{Acknowledgements}

Both S. N. and X. M. Z. would like to acknowledge the hospitality of the department of Physics and Astronomy at UCLA, where this work was initiated. S. N. would like to acknowledge the support of  USA-Israel Binational and Israeli Academy Grants. X. M. Z. thanks Z. H. Lin for discussions. The 
work of R. D. P. was supported in part by the Department of Energy under contract No. DE-FG03-91ER40662, Task C.

\section*{Figure Captions}
\begin{description}
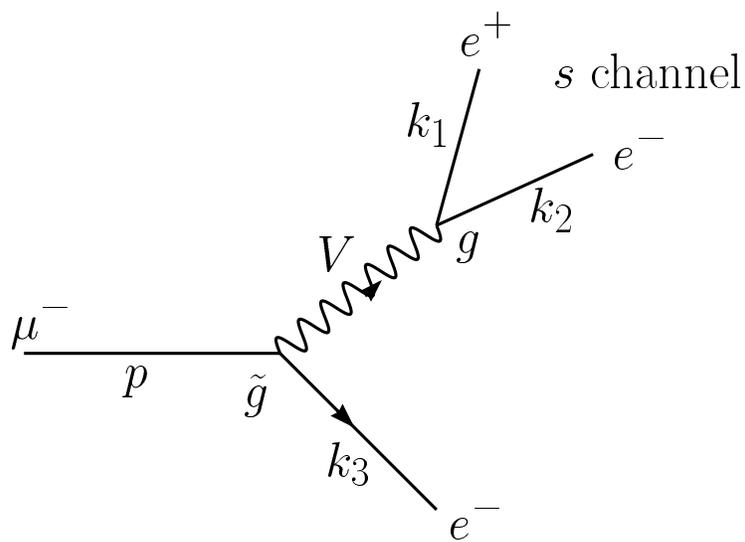
\item{Fig. 1.} A vector exchange diagram contributing to $\mu\to 3e$.
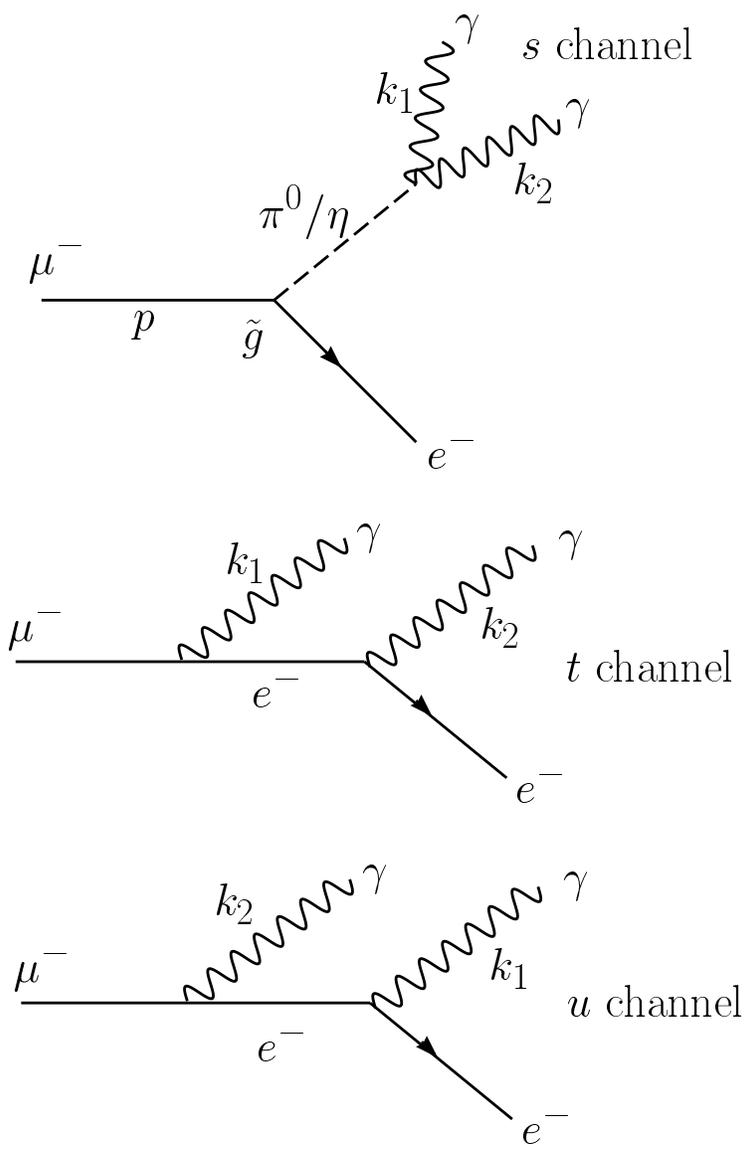
\item{Fig. 2.} The $\pi/\eta^o$ ($s$ channel) and $e$ ($t$ and $u$ channel)
exchange contributing to $\mu\to e\gamma\gamma$.
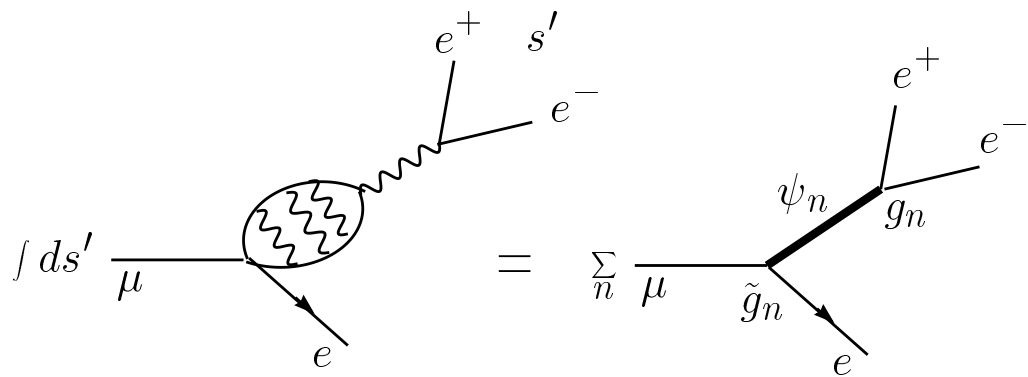
\item{Fig. 3.} The $\bar cc$ bubble and its equivalent description in terms of
$\psi,\psi^{\prime},\ldots,\psi^n$ exchanges.
\end{description}


\section*{References}

\begin{thebibliography}{9}
\bibitem{PDG} Particle Data Group, C. Caso {\it et al.}, Europ. Phys.J. {\bf C3}, 1 (1998).

\bibitem{Kmumu} E871 Collaboration,  D. Ambrose {\it et al.}, Phys. Rev. Lett. {\bf 81}, 4309 (1998).

\bibitem{IJR} J. Illana,
M. Jack, and T. Riemann, hep-ph/0001273, and references therein.

\bibitem{wang} "Probing lepton flavor violation in decays of charmonium and bottomonium systems", R. D. Peccei, J. X. Wang and X. M. Zhang, May 1998 note (unpublished).

\bibitem{SuperK} SuperKamiokande Collaboration, Y Fukuda {\it et al.}, Phys. Rev. Lett. {\bf 81}, 1562 (1998).


\bibitem{DT} See, for example,
P. Langacker and D. London, Phys. Rev. D{\bf 38}, 886 (1988);  D. Tommasini {\it et al.}, Nucl. Phys. B{\bf 444}, 451 (1995).

\end{thebibliography}
\end{document}